\documentclass[12pt]{iopart}
\usepackage{epsf}
\begin{document}

\title {Three-Dimensional FDTD Simulation of Biomaterial Exposure to 
Electromagnetic Nanopulses}

\author{Neven Simicevic \footnote[3]{Correspondence should be addressed to Louisiana Tech University, 
PO Box 10348, Ruston, LA 71272, Tel: +1.318.257.3591, Fax: +1.318.257.4228, 
E-mail: neven@phys.latech.edu}}

\address{\ Center for Applied Physics Studies, Louisiana Tech University,
 Ruston, LA 71272, USA}

\begin{abstract}

Ultra-wideband (UWB) electromagnetic pulses of nanosecond duration, or nanopulses, 
have been recently approved 
by the Federal Communications Commission for a number of various applications.
They are also being explored 
for applications in biotechnology and medicine. The simulation
of the propagation of a 
nanopulse through biological matter, previously performed using a
two-dimensional finite difference-time domain method (FDTD), has been 
extended here into a full three-dimensional computation. To account for
the UWB frequency range, a geometrical
resolution of the exposed sample was $0.25 \; mm$, and
the dielectric properties of biological matter were accurately described in 
terms of the Debye model. The results obtained from three-dimensional 
computation support the previously obtained results: 
the electromagnetic field inside a biological tissue depends on the incident 
pulse rise time and width, with increased importance of the rise time 
as the conductivity increases;  no thermal effects are possible 
for the low pulse repetition rates, supported by recent experiments. 
New results show that the dielectric
sample exposed to nanopulses behaves as a dielectric resonator. For a sample 
in a cuvette, we obtained the dominant resonant frequency and the $Q$-factor of 
the resonator.

\end{abstract}

\pacs{87.50.Rr, 87.17.–d, 77.22.Ch, 02.60.–x}


\maketitle

\section{Introduction}

The bioeffects of non-ionizing ultra-wideband (UWB) electromagnetic (EM) pulses
of nanosecond duration, or nanopulses, have not been studied in as much detail 
as the effects of continuous-wave (CW) radiation. Research on the effects
of high intensity EM nanopulses is only a recent 
development in biophysics (Hu {\it et al} 2005, Schoenbach {\it et al} 2004).
While it has been observed that nanopulses are very damaging to electronics, 
their effects on biological material are not very clear (Miller {\it et al} 2002). 
A typical nanopulse has a width of few nanoseconds, a rise time on the order of 100 
picoseconds, an amplitude of up to several hundreds kilovolts, and
a very large frequency bandwidth. UWB pulses,
when applied in radar systems, have a potential for better spatial resolution, 
larger material penetration, and easier target information recovery (Taylor 1995).
They were approved in 2002 by the Federal Communications Commission 
in the U.S. for ``applications such as radar imaging of objects buried under 
the ground or behind walls and short-range, high-speed data transmissions" (FCC 2002).
It is, therefore, very important to understand their interaction with 
biological materials.

Experiments which can provide a basis for nanopulse exposure safety standards 
consist of exposing biological systems to UWB radiation. The basic exposure 
equipment consists of a pulse generator, an exposure
chamber, such as a gigahertz transverse electromagnetic mode cell (GTEM), and  
measuring instruments (Miller {\it et al} 2002). A typical nanopulse is 
fed into the GTEM cell and propagates virtually unperturbed to the position 
of the sample.
While the pulse generator output can be easily measured, 
the electric field in an exposure chamber in the 
vicinity and inside the sample is difficult or even impossible to measure.
To find the field inside the sample it is necessary to consider a computational 
approach consisting of numerical solution of Maxwell's equations. 

A computational approach requires a realistic description of the geometry 
and the physical properties of exposed biological material, must
be able to deal with a broadband response, and be numerically robust
and appropriate for the computer technology of today. The numerical method
based on the finite difference-time domain (FDTD) method satisfies these conditions.  
This method is originally introduced by Kane Yee in the 1960s (Yee 1966), 
but was extensively developed in the 1990s (Sadiku 1992, Kunz and Luebbers 1993, 
Sullivan 2000, Taflove and Hagness 2000).
 
In the previous paper (Simicevic and Haynie 2005), we applied the FDTD method 
to calculate the EM field inside biological samples exposed to nanopulses in a GTEM cell. 
While the physical properties of 
the environment were included in the calculation to the fullest extent, we 
restricted ourselves to two-dimensional geometry in order to reduce the 
computational time. In this paper we report the results of a full three-dimensional 
calculation of the same problem. We will show that the essential features of the 
two-dimensional solution, such as the importance of the rise time, remain, and that 
full three-dimensional computation produces new results and reveals the 
complexity of the EM fields inside the exposed sample. Since it is possible 
that the bioeffects of short EM pulses are qualitatively different from those 
of narrow-band radio frequencies, knowing all the field components inside
the sample is essential for the 
development of a model of the biological cell, cellular environment, and EM 
interaction mechanisms and their effects (Polk and Postow 1995).

\section{Computational Inputs}

The results presented in this paper are obtained using a full three-dimensional 
calculation of the same FDTD computer code
described in the previous work. This code was 
validated by comparing the numerical and analytical solution of 
Maxwell's equations for a problem which had 
comparable geometrical and physical complexity to the one being 
studied in this work (Simicevic and Haynie 2005). The key requirements 
imposed on the code
is that the space discretization of the geometry and description of the physical 
properties are accurate in the high frequency domain associated with an UWB pulse
and appropriate for numerical simulation. 

\begin{figure}
\begin{center}
\mbox{\epsfxsize=3.5in\epsfbox{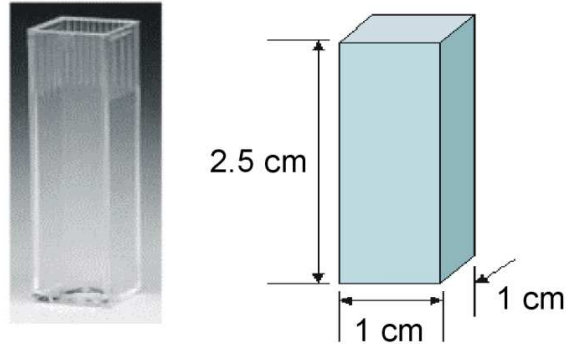}}
\vspace*{- 0.5cm}
\caption{\label{expo} Shape and size of the polystyrene cuvette used in 
the computation.}
\end{center}
\end{figure}

The computation consists of calculating EM fields inside the polystyrene
cuvette, shown in Figure~\ref{expo}, filled with biological material 
and exposed to UWB radiation.
The size of the cuvette is $1 \; cm$ $\times$ $1 \; cm$ $\times$ $2.5 \; cm$, 
with $1 \; mm$ thick walls. In order to compare the results 
from this work and the results from the previous two-dimensional calculation, 
the material inside the cuvette was the same, blood or water.

\begin{figure}
\begin{center}
\mbox{\epsfxsize=3.5in\epsffile{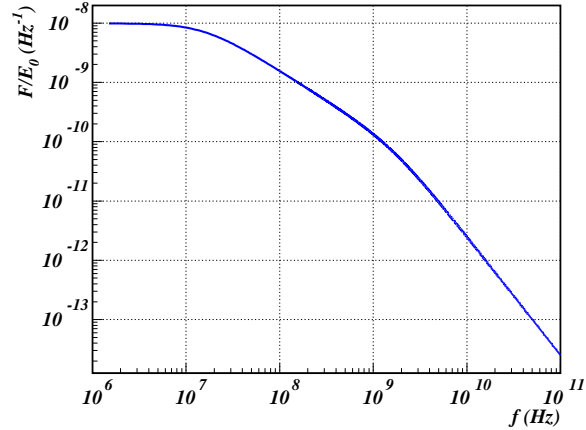}}
\vspace*{- 1.5cm}
\end{center}
\caption{Frequency spectrum of the double exponential pulse, described in
Equation~\ref{dubexp}, at the position of the source. $F/E_{0}$ denotes Fourier
transform normalized to the field amplitude $E_{0}$.}
\label{four}
\end{figure}

The cuvette was exposed to a vertically polarized EM pulse the shape of 
which is described as a double exponential function (Samn and Mathur 1999)

\begin{equation} E=E_{0}(e^{-\alpha t}-e^{-\beta t}).
\label{dubexp}
\end{equation}
$E_{0}$ is pulse amplitude and coefficients $\alpha$ and $\beta$ 
define the pulse rise time, fall time, and width. 
Numerical values of the parameters are roughly the ones measured in the
GTEM cell used for bioelectromagnetic research at Louisiana Tech University:
$E_{0} = 18.5 \; kV/m$, $\alpha = 1.0 \times 10^{8} \; s^{-1}$, 
and $\beta = 2.0 \times 10^{10} \; s^{-1}$. This pulse has a rise time of  
$\sim 150 \; ps$ and a width of $\sim 10 \; ns$. While detailed properties of
a double exponential pulse can be found elsewhere (Dvorak and Dudley 1995),
for better understanding of the results presented in this paper it is useful
to know the frequency spectrum of the pulse used, which we plotted in 
Figure~\ref{four}.   

The shape of the cuvette and the sample inside was discretized
by means of Yee cells, cubes of edge length $\Delta x$. 
The Yee cells had to be small enough not to distort the shape
and large enough for the time step, calculated from the Courant stability criterion 
(Taflove and Brodwin 1975, Kunz and Luebbers 1993, Taflove and Hagness 2000)

\begin{equation}
\Delta t \leq {1 \over {c \sqrt{(\Delta x)^{-2}+(\Delta y)^{-2}+(\Delta z)^{-2}}}},
\label{Cour}
\end{equation}
to be practical for overall computation. In Equation \ref{Cour}, in our case, 
$\Delta x = \Delta y = \Delta z$ and $c$ is the speed of light in vacuum. 
In order for the computation to be appropriate for the full frequency range of 
the EM pulse, the size of a Yee cell must also satisfy the rule

\begin{equation}
\Delta x \simeq {v \over {10 \; f_{max}}},
\end{equation}
where $v$ is the speed of light in the material and $f_{max}$ is the 
highest frequency considered defined by the pulse rise time
\begin{equation}
f_{r}={0.35 \over \tau_{r}}.
 \label{Rtime}
\end{equation}
In this equation $f_{r}$ is a maximum frequency in $Hz$ and $\tau_{r}$ is a 
rise time in $s$ (Faulkner 1969).

In the previous, as well as in the present work, the Yee cube edge length of 
$\Delta x = 1/4 \; mm$ satisfies all the above 
criteria and may be used to model 
blood exposure to the wave frequency of up to $ 15 \; GHz$, a
much greater value than required by the rise time criterion.
The time step derived from the Courant criterion is $\Delta t \leq 0.48 \; ps$.
Optimal agreement between geometrical and physical descriptions eliminated 
the need for additional approximations and resulted in about one hour 
of computational time for
every nanosecond of simulated time on a modern personal computer.

\section{Dielectric Properties of Exposed Sample}

Proper description of the dielectric properties of the exposed material
is crucial when dealing with an UWB electromagnetic pulse.
EM properties of a biological material are normally expressed in terms 
of frequency-dependent dielectric properties and conductivity, usually
parametrized using the Cole-Cole model (Gabriel 1996, Gabriel {\it et al} 1996):

\begin{equation}
\varepsilon(\omega) = \varepsilon_{\infty} + \sum_{k=1}^{4}
{\Delta \varepsilon_{k} \over {1+(i\omega\tau_{k})^{1-\alpha_{k}}}} 
+ {\sigma  \over i\omega \varepsilon_{0}},
\label{CC}
\end{equation}
where $i=\sqrt{-1}$, $\varepsilon_{\infty}$ is the permittivity in the 
terahertz frequency range, $\Delta \varepsilon_{k}$ are the drops in 
permittivity in a specified frequency range, $\tau_{k}$ are the relaxation 
times, $\sigma$ is the ionic conductivity, and  $\alpha_{k}$ are the coefficients
of the  Cole-Cole model. They constitute up to 14 real parameters of a 
fitting procedure. 
While this function can be numerically Fourier transformed into the time domain, 
its application is problematic for FDTD. In addition to the physical problems
arising when the Cole-Cole parametrization is applied (Simicevic and Haynie 2005), 
this parametrization requires time consuming numerical
integration techniques and makes computation unacceptably slow.

If instead of a Cole-Cole parametrization one uses the Debye model in which
the dielectric properties of a material are described as a sum of $N$ 
independent first-order processes

\begin{equation}
\varepsilon(\omega) = \varepsilon_{\infty} + \sum_{k=1}^{N}
{ \Delta \varepsilon_{k} \over {1+i\omega\tau_{k}}}=\varepsilon_{\infty} 
+\sum_{k=1}^{N} \chi_{k}(\omega),
\label{Dpar}
\end{equation}
then for each independent first-order process the Fourier transformation 
has an analytical solution

\begin{equation}
\chi_{k}(t)={ \Delta \varepsilon_{k} \over \tau_{k}} \; e^{-t/\tau_{k}}, 
\; \;  {t \geq 0}.
\label{Dpart}
\end{equation}
$\chi(\tau)$ is the electric susceptibility of a material (Jackson 1999) and
 $\tau_{k}$ is the relaxation time for process $k$.

The static conductivity, $\sigma$, is defined in the time domain as 
the constant of proportionality between the current density and 
the applied electric field, $\vec J = \sigma \vec E$, and its implementation 
in FDTD does not require additional or different Fourier transforms 
(Kunz and Luebbers 1993).  

The Debye parametrization allows use of  a recursive 
convolution scheme (Luebbers {\it et al} 1990, 1991, Luebbers and Hunsberger 1992) 
and makes FDTD computation 
an order of magnitude faster compared to the use of numerical integration. In 
a recursive convolution scheme the permittivity at time step $(m+1)$ is simply 
the permittivity at time step $m$ multiplied by a constant (Kunz and Luebbers 1993). 
In addition to making the computation faster, in the previous work 
(Simicevic and Haynie 2005) we have shown that the Debye model 
also provides a sufficiently accurate description of 
physical properties of some biological materials.  Here we use the 
same Debye parameters used in the previous two-dimensional calculation.
The parameters applied in the Debye model of the form

\begin{equation}
\varepsilon(\omega) = \varepsilon_{\infty} 
+{{\varepsilon_{s1}-\varepsilon_{\infty}} \over {1+i\omega\tau_{1}}}
+{{\varepsilon_{s2}-\varepsilon_{\infty}} \over {1+i\omega\tau_{2}}}
\label{Dpar2}
\end{equation}
are shown in Table ~\ref{tab1}.

\begin{table}
\begin{center}
\begin{tabular}{lcccccc}
\multicolumn{1}{c}{Material} {\vline}&
\multicolumn{1}{c}{$\varepsilon_{\infty}$} {\vline}&
\multicolumn{1}{c}{$\varepsilon_{s1}$} {\vline}&
\multicolumn{1}{c}{$\varepsilon_{s2}$} {\vline}&
\multicolumn{1}{c}{$\tau_{1} (s)$} {\vline}&
\multicolumn{1}{c}{$\tau_{2} (s)$} {\vline}&
\multicolumn{1}{c}{$\sigma (S/m)$} {\vline}\\
\hline\hline
   Polystyrene            &  2.0  &  -     &  -   &  - & - &0.\\
   Water              &  4.9  &  80.1  &  -   & $10.0 \; 10^{-12}$ & - &0.\\
   Blood              &  6.2  & 2506.2 & 65.2 &  $9.0 \; 10^{-8}$  & $8.37 \; 10^{-12}$   &0.7\\
\end{tabular}
\end{center}
\caption{Debye parameters for the materials used in the computation 
(Simicevic and Haynie 2005).}
\label{tab1}
\end{table}

\section{Field Calculation and Data Representation}

FDTD calculations of the exposure of a biological material to EM nanopulses provide 
the values of electric and magnetic field components at every space point 
and throughout the time range. In the case of a three-dimensional 
computation, there is 
overwhelming information such that  data reduction and representation of the results 
becomes a nontrivial task. Contrary to radar applications where the interest is
in scattered fields, here we care about total fields inside 
and closely sourrounding the sample. 
Even in such a restricted volume we have an immense number of data points and one
has to carefully select the region of interest and the information to collect.
The extraction of data depends on the physical model of interest and has to be decided
prior to running the FDTD program.
  
\begin{figure}
\begin{center}
\mbox{\epsfxsize=3.in\epsffile{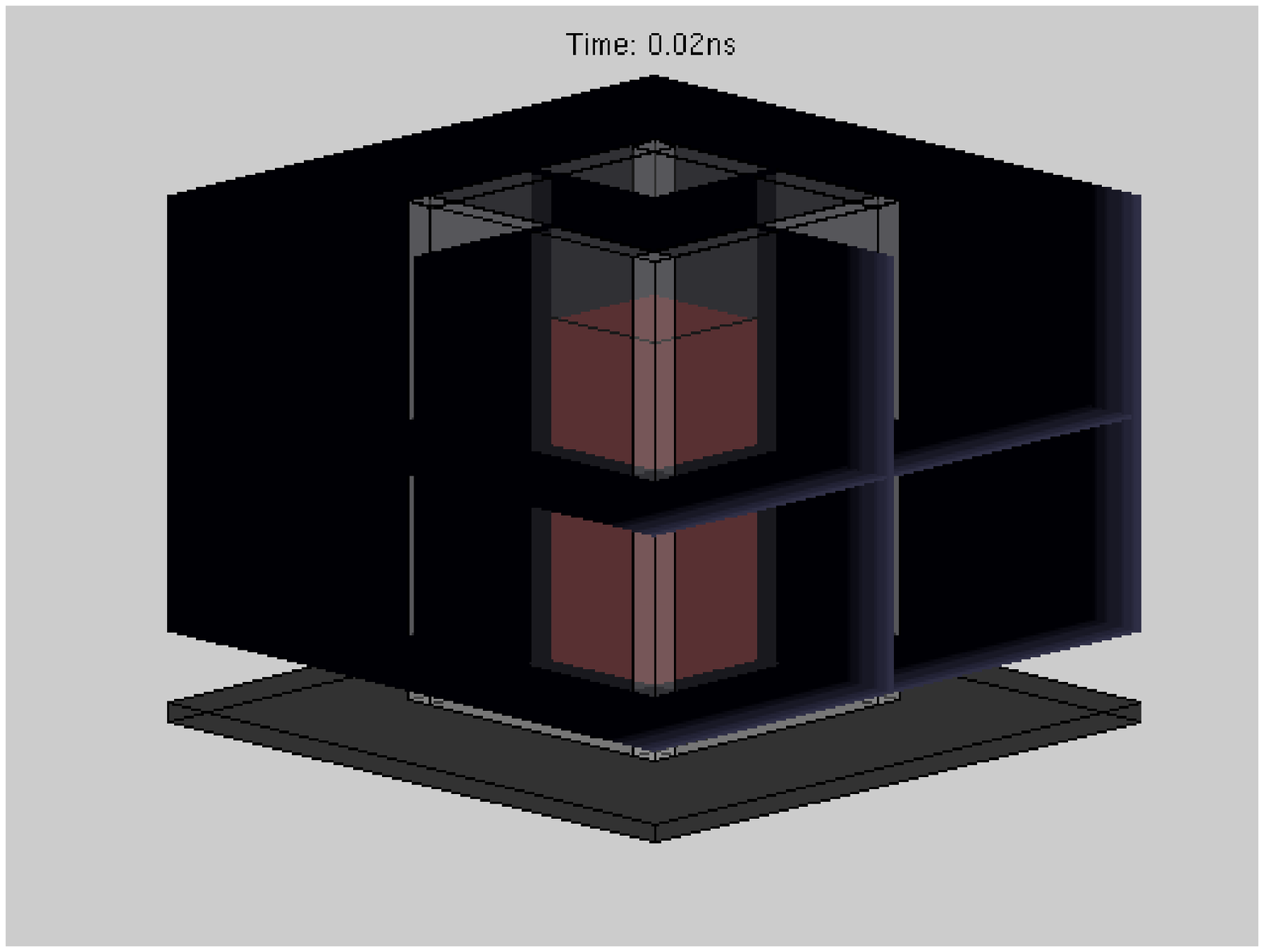}
\epsfxsize=3.in\epsffile{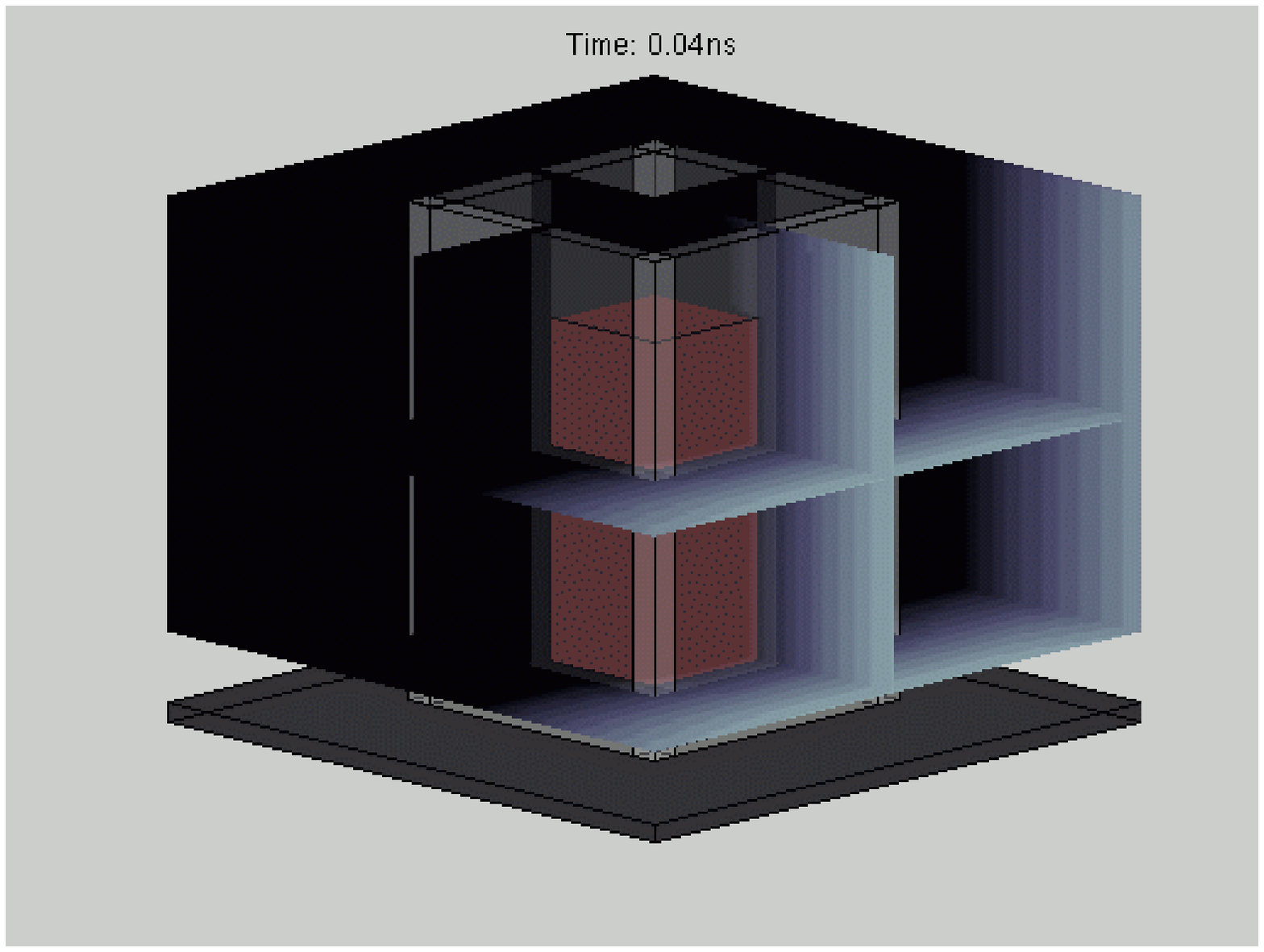}}
\mbox{\vspace*{-5.8cm}\epsfxsize=3.in\epsffile{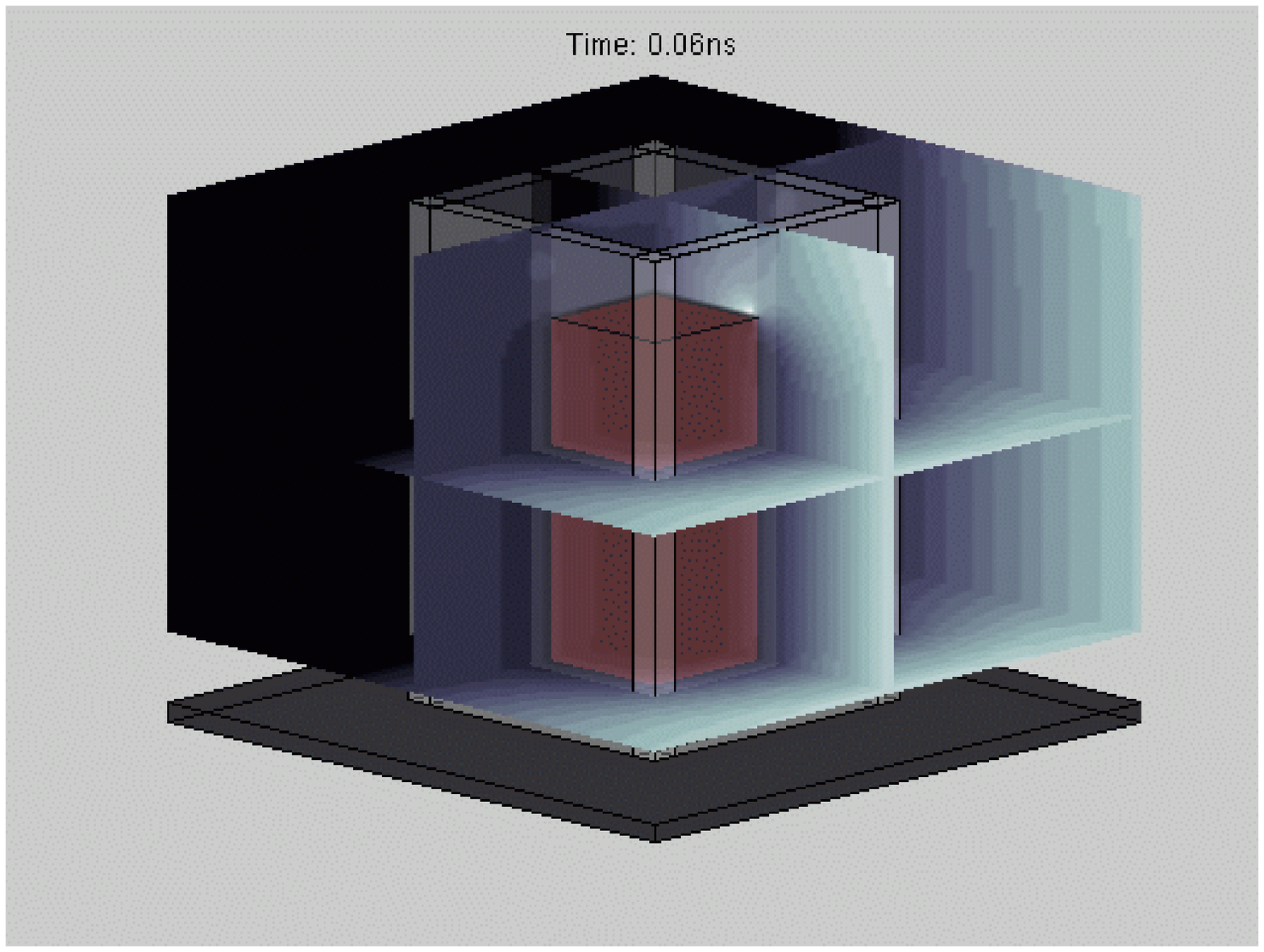}
\epsfxsize=3.in\epsffile{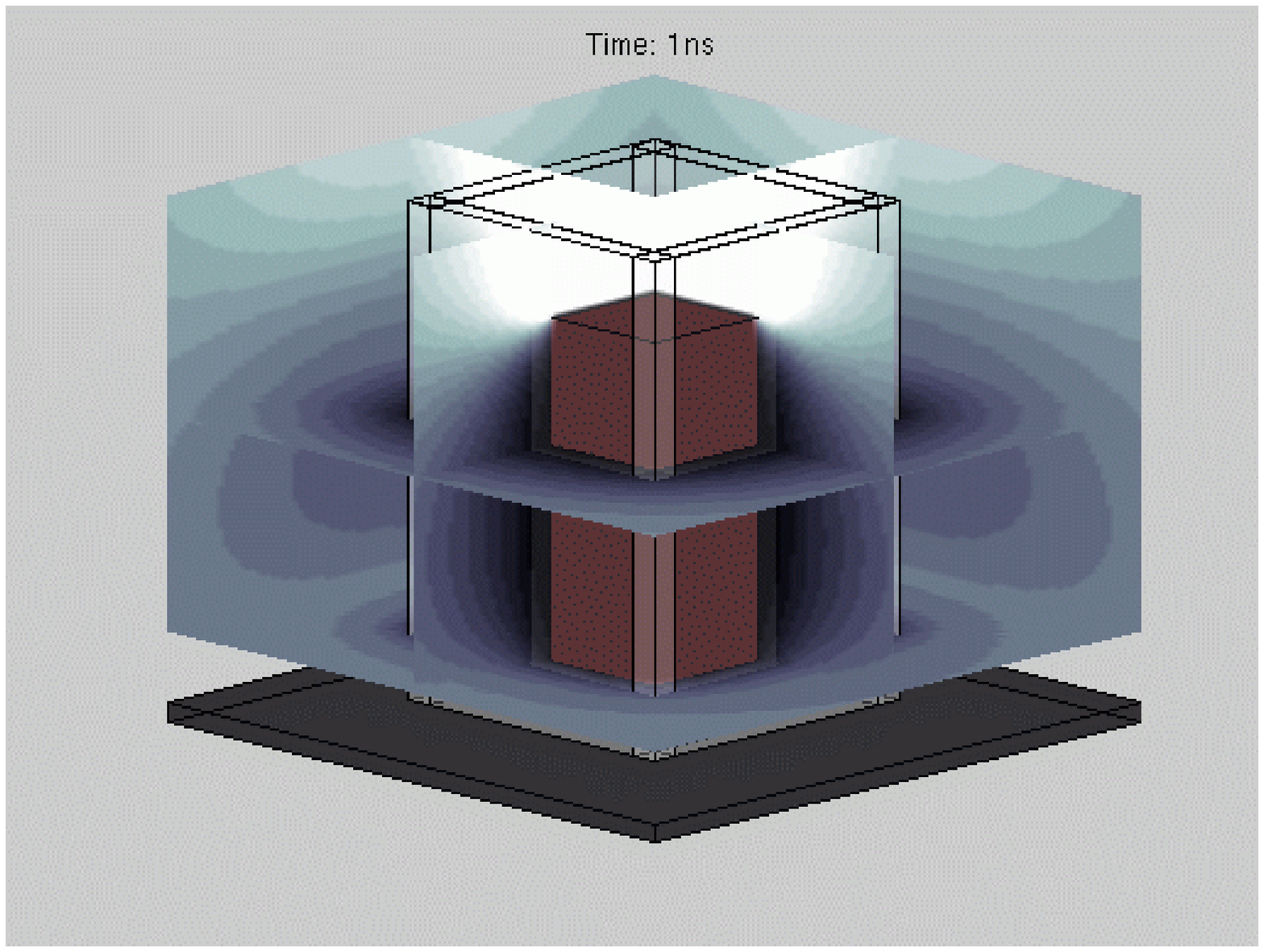}}
\end{center}
\caption{Penetration of an EM pulse into a blood-filled polystyrene cuvette.  
The contours represent the $z$-component of the electric field in steps of 1000 $V/m$.
The dimension of the active area is $3 \; cm$ $\times$ $3 \; cm$ $\times$ $3 \; cm$.
Brighter regions correspond to a stronger field.} 
\label{snaps}
\end{figure}

The FDTD method enables easy creation of animated 
movies, which are very useful as a first step in the analysis and understanding 
of the behavior of the EM fields in space and time. 
The restriction of such visualization is that only parts of the full result
can be represented at any given time and only in a chosen region of interest,
typically in a few selected planes.
As an example, snapshots of the penetration of an EM pulse into the
cuvette, shown in Figure \ref{expo}, filled with blood to the height of $ 2 \; cm$  
are shown in Figure \ref{snaps}. The complete animation can be accessed 
on-line (Simicevic 2005).

\section{Results}

In the previous two-dimensional calculation we have shown that the penetration of 
a linearly-polarized EM pulse
described by Equation~\ref{dubexp} into the blood-filled cuvette
is governed by the pulse rise-time, creates a sub-nanosecond pulse, and 
is absorbed into a conductive loss of the material. 
While confirming the same results, the three-dimensional calculation 
also shows that the blood-filled 
cuvette behaves as a rectangular dielectric resonator. 

In general, if a dielectric object is immersed in a incident sinusoidal wave, 
the EM fields in and around the object peak to high values at certain resonant 
frequencies. The object has a property of a resonator. The properties of 
dielectric resonators, such as 
resonant frequencies, field patterns, and quality factors, are difficult 
to obtain analytically except for a simple shapes such as, for example, a sphere
(Van Bladel 1975). The FDTD method in combination with other 
techniques, like Fourier analysis or Prony's method, can be used to determine the 
resonant frequencies and quality factors of dielectric resonators numerically
(Navarro {\it et al} 1991, Harms {\it et al} 1992, Pereda {\it et al} 1992).  

The resonant frequencies of a dielectric resonator depend on the size, shape 
and dielectric properties of the resonator (Kajfez and Guillon 1986). 
Generally, they can be found from the
boundary condition at the surface between the resonator and the 
surrounding medium (Balanis 1989, Jackson 1999).
In the case of a rectangular cavity resonator, a metal box, which can be filled 
with dielectric material, that traps 
the electromagnetic field, the
resonant frequencies can be easily calculated using (Harrington 1961)

\begin{equation}
f_{mnp} ={ 1 \over {2 \sqrt{\varepsilon\mu}}} 
\sqrt{\left({m \over a}\right)^{2}+\left({n \over b}\right)^{2}
+\left({p \over c}\right)^{2}},
\label{fmnp}
\end{equation}
where $m=0,1,2,... \;$; $n=0,1,2,...\; $; $p=0,1,2,...\; $ are used as a label 
of the resonant mode. Also, 
$f_{mnp}$ is the corresponding resonant frequency, $\varepsilon$ and $\mu$ 
are the permittivity and permeability, respectively. 
The quantities a, b and c are the dimensions of the 
rectangular resonator.

For a pure dielectric box the situation is more complicated. 
The field is not entirely confined inside the box but 
it also exists as an evanescent wave outside the box. 
This wave decays exponentially with the distance from the 
dielectric. In our case, the complications arise also from the constant change 
of the exterior field 
caused by the incident pulse. The boundary conditions require that any tangential
component of the electric field $\vec E$ be continuous, and that any normal 
component of electric displacement $\vec D$ be discontinuous by the 
amount of the charge density on the surface. Also, any normal component of
the magnetic flux density $\vec B$ has to be continuous, and any tangential
component of the magnetic field $\vec H$ has to be discontinuous by the 
amount of the surface current density. More detailed discussion 
of the boundary conditions and resonant modes of the dielectric resonator 
can be found in the paper by R. K. Mongia and A. Ittipiboon and references therein 
(Mongia and Ittipiboon 1997). Application of the boundary conditions in the FDTD
calculation is discussed in more details by P. Yang {\it et al} (Yang {\it et al} 2004).

\begin{figure}
\begin{center}
\mbox{\hspace*{-1.cm}\epsfxsize=4.5in\epsffile{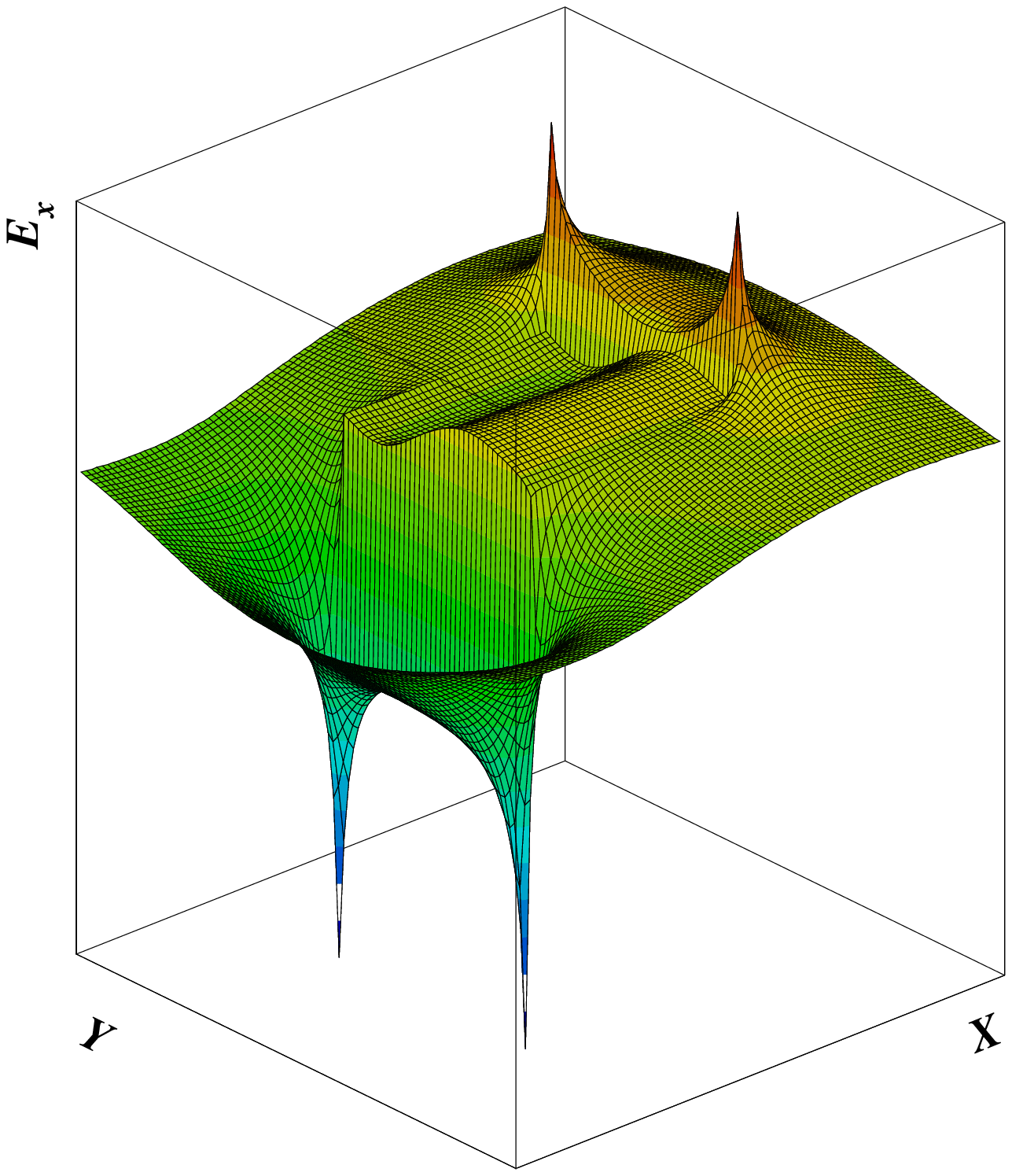}
\hspace*{-4.cm}\epsfxsize=4.5in\epsffile{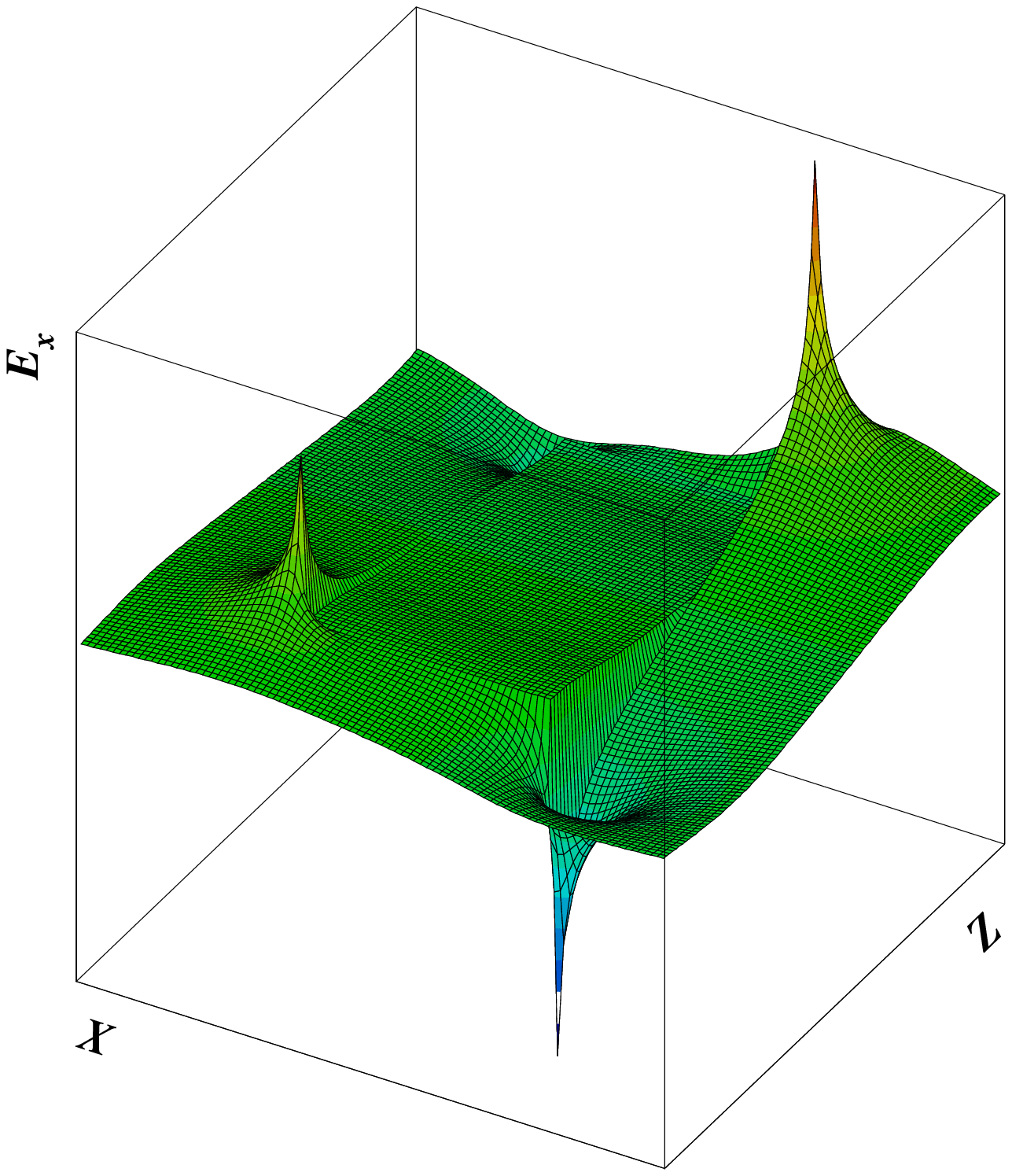}}
\vspace*{- 0.5cm}
\end{center}
\caption{The $E_{x}$ component of the electric field inside and around
the dielectric. Left: in the horizontal X-Y plane
across the midpoint of the cuvette. Right: in the vertical X-Z plane 
across the same point. To understand the resonant behavior better, the 
cuvette walls were neglected in those pictures. The field values are taken 
160 $ps$ after the pulse first impact on the dielectric.}
\label{exbox}
\end{figure}

To find out what is happening when a  dielectric  box is 
immersed in the electromagnetic pulse,
we have plotted the values of the x-component of the total field, $E_{x}$,
for a selected time and in two planes: in the horizontal X-Y plane
across the midpoint of the cuvette (left side of the Figure \ref{exbox}), and 
in the vertical X-Z plane across the same point (right side of the 
Figure \ref{exbox}). As a result of the boundary conditions, in both cases 
$E_{x}$ is discontinuous at the 
wall normal to its direction. At the wall parallel to the field direction, 
$E_{x}$ has to be continuous, but, since the outside field has 
to be continuous too, there is an abrupt change of
the value of the $E_{x}$ component along the parallel wall, a change that is 
more prominent closer to the edges of the wall. While the details of the
field behavior change from one time point to an other, the general 
feature stays the same. 
It is important to notice the formation of a resonant wave inside the 
dielectric on both sides of Figure \ref{exbox}.

\begin{figure}
\begin{center}
\mbox{\epsfxsize=4.in\epsfbox{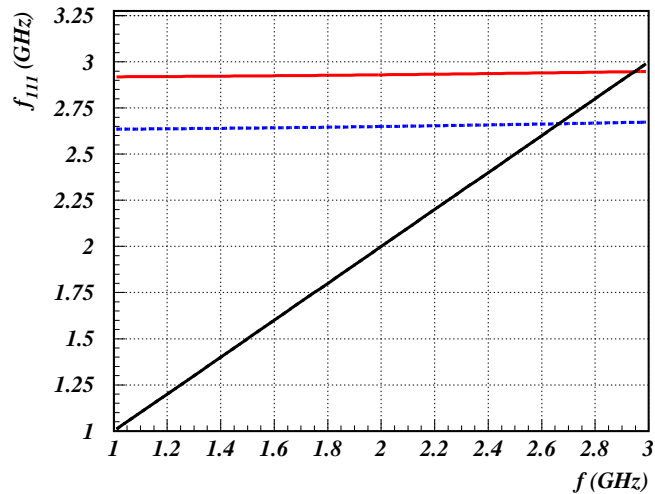}}
\vspace*{- 1.5cm}
\caption{\label{fmnp_f} The resonant frequency as a function of frequency for 
blood (higher full red horizontal line) and water (lower dotted blue line).
The resonant frequencies are shown only in the frequency region of interst. 
Intersections of the function  $f_{111}=f$ with those lines are the
expected resonant frequencies of the dielectric resonator.}
\end{center}
\end{figure}

It is not easy to estimate the resonant frequencies of a dielectric slab 
(Antar {\it et al} 1998). In the first approximation, assuming that 
the exponential decay
of the evanescent wave with the distance from the slab is very fast, one can 
estimate the resonant frequency using Equation \ref{fmnp}. We have calculated 
the lowest resonant frequency by selecting $m=n=p=1$, which allows for all the 
components of the EM field to exist.
In our case $a=b=0.95 \; cm$ and $c=2.0 \; cm$. Since $\varepsilon$ in 
Equation \ref{fmnp} is a function of 
frequency described by the Debye model, the resonant frequency will be a function of 
frequency, too. If the dielectric resonator were immersed in the plane wave, 
we would expect the resonance to occur when the frequency of the wave is equal
to the resonant frequency. In Figure \ref{fmnp_f} this corresponds to the
intersection of the line $f_{111}=f$ and calculated resonant frequencies.
For water the expected resonant frequency was found to be $f_{111}= 2.66 \; GHz$ 
and for blood $f_{111}= 2.94 \; GHz$. As shown in Figure \ref{four}, those
frequencies are well inside the frequency spectrum of the 
double exponential pulse used in this calculation, therefore one can expect the 
resonant excitation to occur.

\begin{figure}
\begin{center}
\mbox{\epsfxsize=4.in\epsfbox{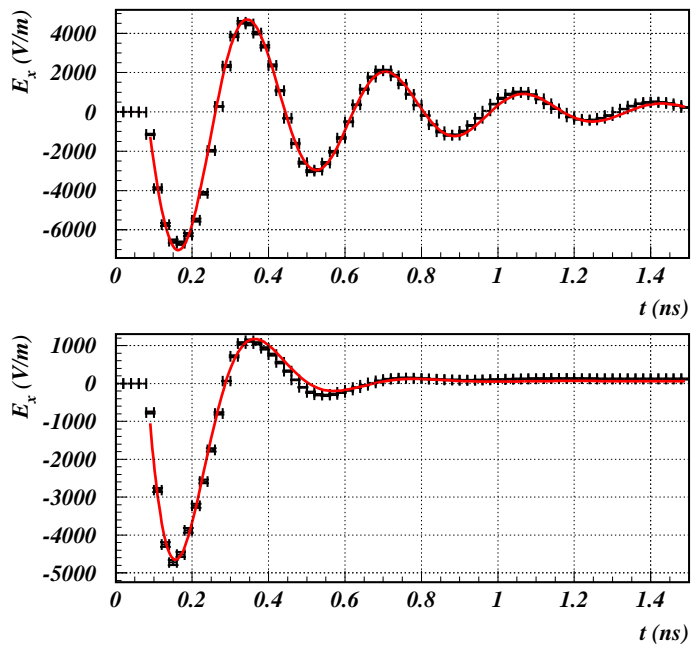}}
\vspace*{- 0.5cm}
\caption{\label{res_wb} FDTD data at the cuvette's mid point as a function of time.
The upper histogram shows the data for water, and lower histogram for blood. 
The curves are the fits using Equation \ref{damp_low}.} 
\end{center}
\end{figure}

To estimate the resonant frequencies using the FDTD data the
field values were extracted at a fixed observation point inside the 
dielectric resonator as  
function of time. For any field component those values can be expressed 
in the form of free damped oscillator (Ko and Mittra 1991, Pereda {\it et al} 1992). 
Assuming just one resonant mode and selecting the $E_{x}$ component of the
total field one can write

\begin{equation}
E_{x}(t) = A e^{-\alpha t} \sin(2\pi f_{r} t) + B,
\label{damp_low}
\end{equation}
where $A$ is the modal amplitude, $\alpha$ is the damping factor, $f_{r}$ is 
the resonant frequency, and $B$ is small additional noise. The FDTD data 
fitted with this function are shown in Figure \ref{res_wb}. The results
of the fit are tabulated in Table ~\ref{tab2}. The ratio $B/A$ is less than 1\%. 
Taking into consideration the simplicity of our model, the agreement between the 
expected resonant freqency and the one obtained through the fit is very good.

Knowing the
resonant frequency and the damping factor, one can also obtain the quality factor 
$Q$ of the dielectric resonator using
\begin{equation}
Q = \pi f_{r}/ \alpha.
\label{q_fact}
\end{equation}
The $Q$-factor is also tabulated in Table ~\ref{tab2}.

\begin{table}
\begin{center}
\begin{tabular}{lcccccc}
\multicolumn{1}{c}{} {\vline}&
\multicolumn{1}{c}{$f_{r}$ (Eq. \ref{fmnp})} {\vline}&
\multicolumn{1}{c}{$f_{r}$ (Eq. \ref{damp_low})} {\vline}&
\multicolumn{1}{c}{$\alpha$} {\vline}&
\multicolumn{1}{c}{$Q$} {\vline}\\
\hline\hline
 Water  & $2.66 \; GHz$ & $2.77 \; GHz$ & $2.35 \; 10^{9}\; s^{-1}$ & $3.7$ \\
 Blood  & $2.94 \; GHz$ & $2.49 \; GHz$ & $6.97 \; 10^{9}\; s^{-1}$ & $1.1$ \\
\end{tabular}
\end{center}
\caption{Estimated and obtained resonant frequencies, damping factors and 
$Q$-factors for water and blood filled cuvette.}
\label{tab2}
\end{table}

\begin{figure}
\begin{center}
\mbox{\epsfxsize=3.5in\epsffile{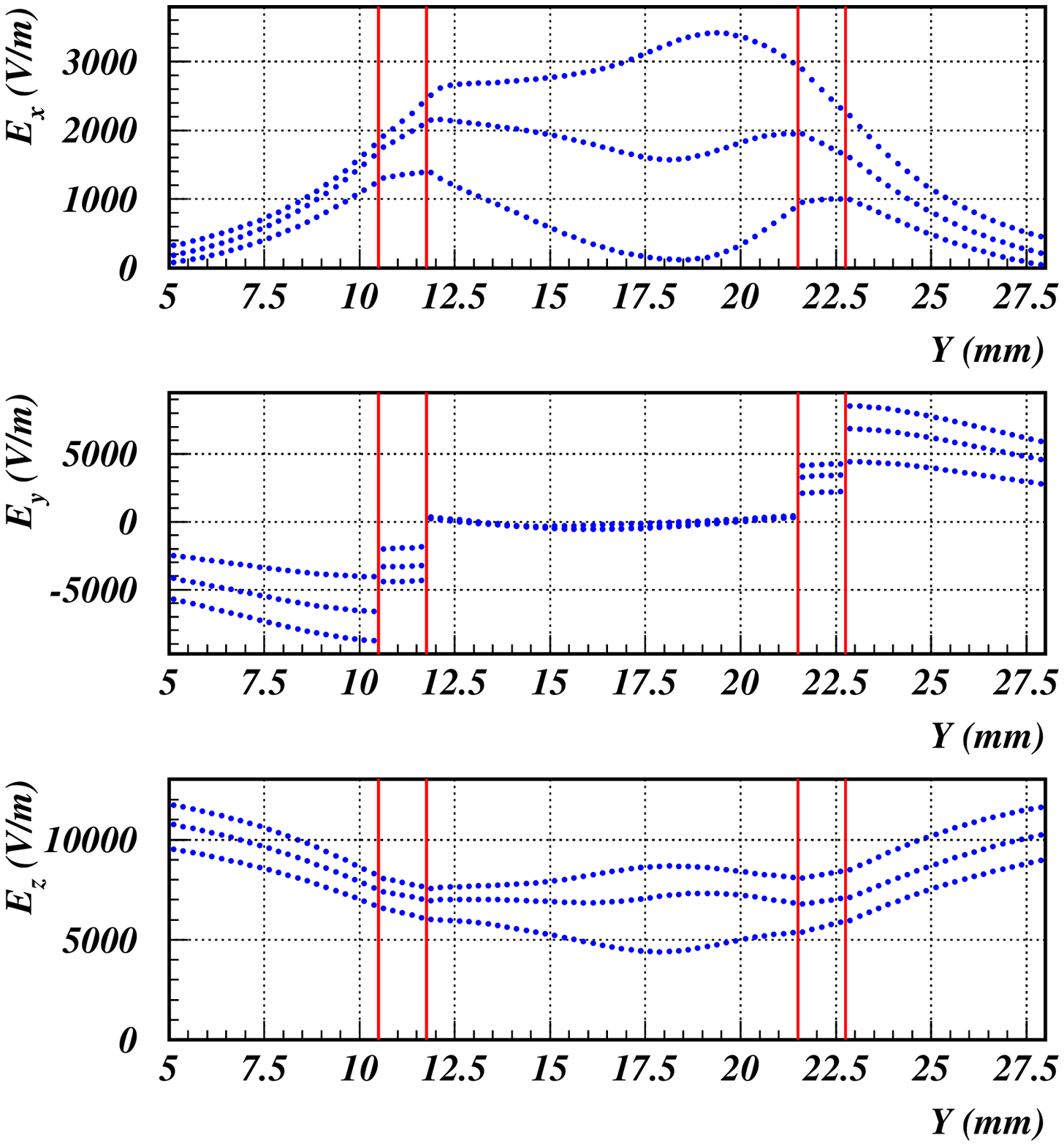}
\hspace*{-1.3cm}\epsfxsize=3.5in\epsffile{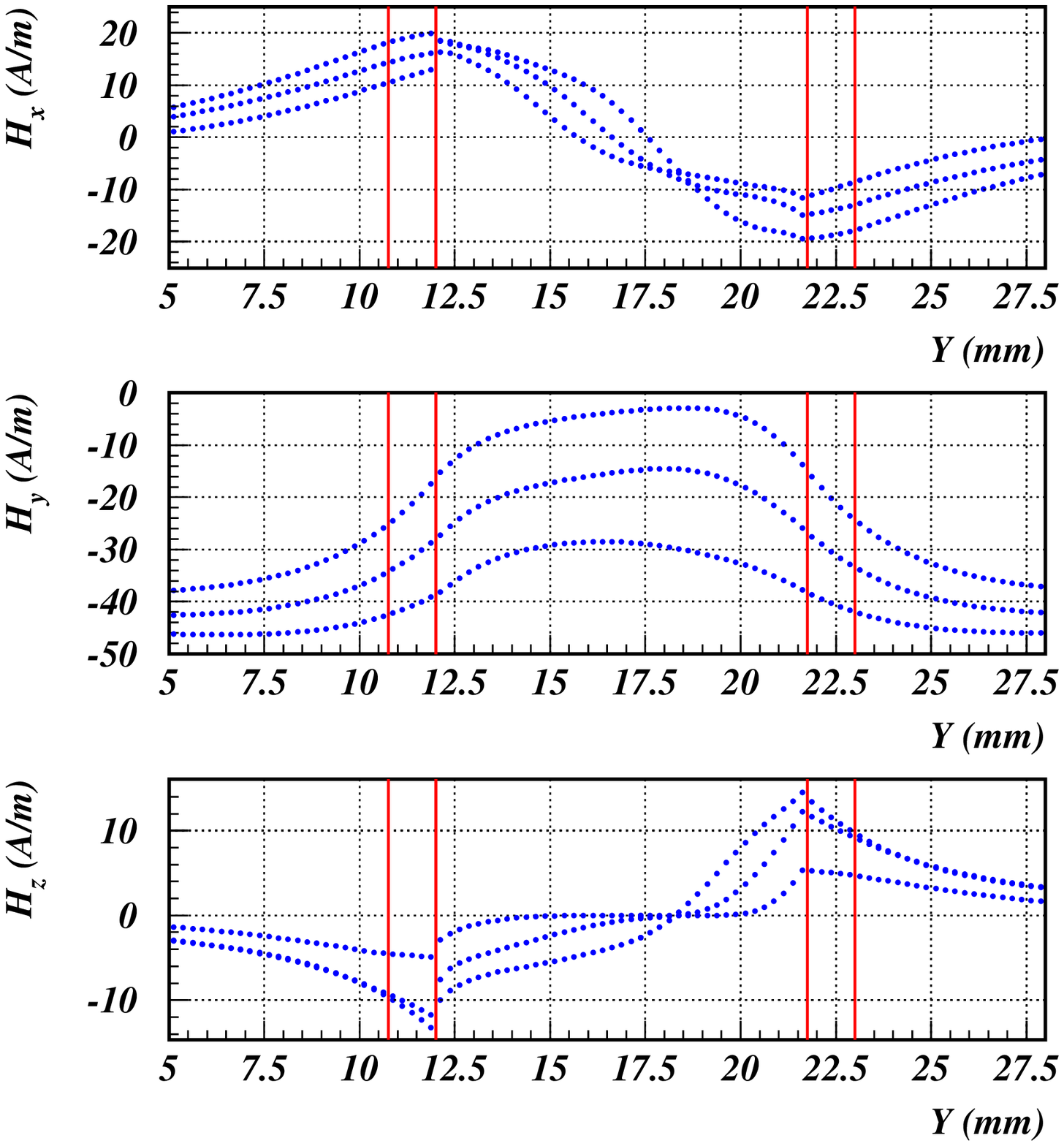}}
\vspace*{- 1.0cm}
\end{center}
\caption{The values of all the electric and magnetic 
field components in the blood and in the cuvette walls, in the x-z plane, 
during the rise time of the pulse for three time periods separated by 20 $ps$.}
\label{ehxyz}
\end{figure}

Finally, Figure \ref{ehxyz} shows a few values of all the electric and magnetic 
field components in the blood and in the cuvette walls in the x-z plane 
during the pulse rise time. The curves are separated in time by 20 $ps$.
As expected, the $E_{x}$, $E_{z}$, and $H_{y}$ are continuous and 
$E_{y}$, $H_{x}$, and $H_{z}$ are discontinuous at the boundaries of the materials.
The buildup of a resonant steady wave is also shown.

The full three-dimensional computation of the exposure of the blood-filled 
cuvette to a linearly-polarized EM pulse 
described by Equation~\ref{dubexp}, not only reproduced the results
of the two-dimensional computation, but also gave an insight into properties of induced
components of the EM field. The three-dimensional calculation supports the notion
that pulse penetration is a function of both rise time and pulse width,
with both pulse features important in the case of a non-conductive material,
right side of Figure \ref{dubexp}, and the
penetration dominated by rise time in the case of a conductive material,
left side of Figure \ref{dubexp}.
It also shows that for a material of considerable conductivity, the incident 
pulse width is relatively unimportant. In addition, the three-dimensional calculation
reveals that the dielectric box exposed to EM pulses behaves as a dielectric
resonator.  
 
\begin{figure}
\begin{center}
\mbox{\epsfxsize=3.in\epsffile{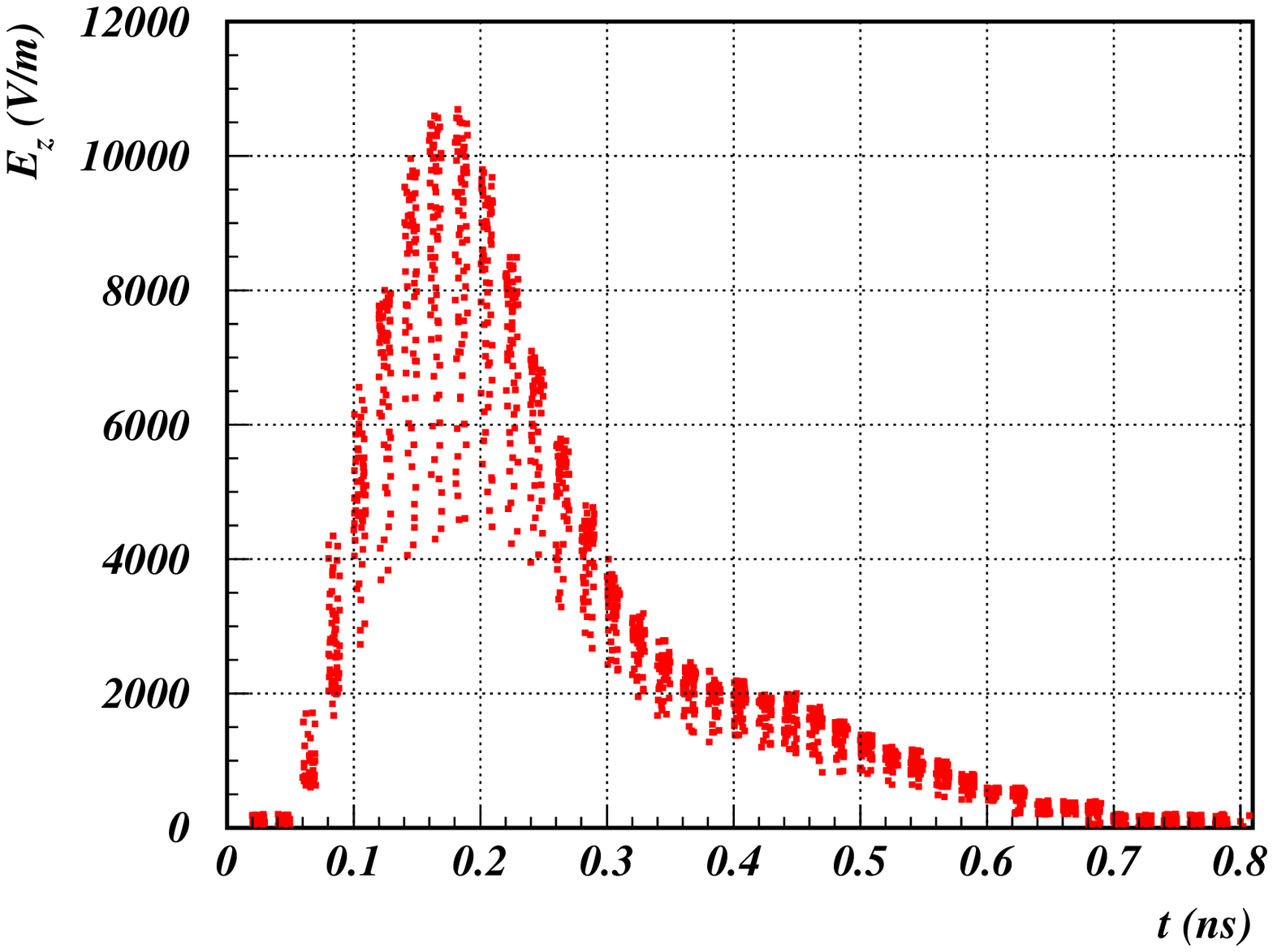}}
\mbox{\vspace*{-5.cm}\epsfxsize=3.in\epsffile{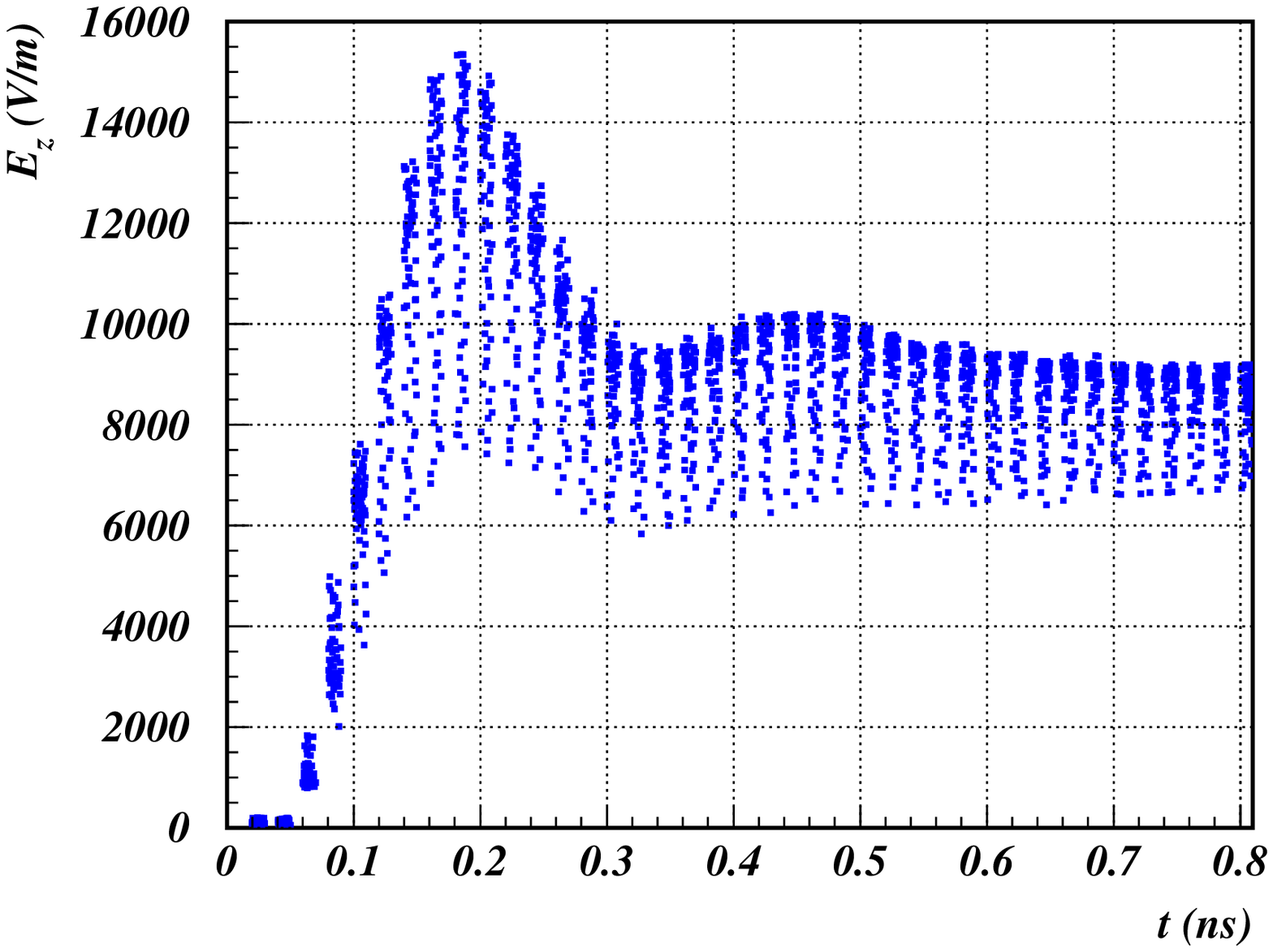}}
\vspace*{- 1.cm}
\end{center}
\caption{Left: the pulse shape of a blood-penetrated field component in 
the direction of pulse polarization for the first 0.8 $ns$. 
Right: the pulse shape of a water-penetrated field component in 
the direction of pulse polarization for the first 0.8 $ns$.
Distribution of the field values in a particular time interval is a measure 
of the inhomogeneity of the field across the sample.}
\label{eblood_water}
\end{figure}

FDTD approach also allows calculation of the energy deposited in a 
biological material using (Jackson 1999)

\begin{equation}
{{dE} \over {dt}}= \int_{V} \vec J \cdot \vec E  \;dV  \;=  \;
\sigma \; \int_{V} \vec E \cdot \vec E  \;dV,
  \label{power}
\end{equation}
where ${dE} \over {dt}$ is the total energy per unit time absorbed in the 
volume, $\vec J$ and $\vec E$ are, respectively, the current density and 
electric field inside the material, and $\sigma$ is the conductivity of the
material. 
FDTD computation provides all the above values through the entire time and it is easy to
perform numerical integration of Equation \ref{power}.   
The results from the three-dimensional calculation 
are the same as from the two-dimensional calculation. For blood,
the average converted energy per pulse, 
described by Equation~\ref{dubexp}, is again $\sim 0.003 \; J/m^{3}$.  
The resulting temperature increase of
$\sim 10^{-10} \; K$ per pulse is clearly negligible if the pulse repetition rate 
is low.  Experiments performed by P. T. Vernier {\it et al}
(Vernier {\it et al} 2004) have shown that even the nanopulses
of similar duration but orders of magnitude higher field amplitude 
than the ones used in this work do not not cause 
significant intra- or extra-cellular Joule heating in the case of a low 
pulse repetition rate.

\section{Conclusion}

In this paper we have extended our previous FDTD calculations on nanopulse 
penetration into biological matter from two to three dimensions.  
Calculations included the same detailed geometrical description of the material 
exposed to nanopulses, the same accurate description 
of the physical properties of the material, the same spatial resolution
of $1/4 \; mm$ side length of the Yee cell, and the same cut-off frequency 
of $\sim 100 \; GHz$ in vacuum and $\sim 15 \; GHz$ in the dielectric. 
To minimize computation time, the dielectric 
properties of a tissue in the frequency range $\leq 100 \; GHz$ were formulated  
in terms of the Debye parametrization which we have shown in a previous paper 
to be, for the materials studied, as accurate as the Cole-Cole parametrization.

The results of three-dimensional FDTD calculation can be summarized 
as follows:

a) The shape of a nanopulse inside a biomaterial is a function of both rise time 
and width of the incident pulse, with the importance of the rise time increasing 
as the conductivity of the material increases. Biological cells inside 
a conductive material are exposed to pulses 
which are often substantially shorter than the duration of 
the incident pulse. The same results followed from the two-dimensional calculation.

b) The dielectric material exposed to the EM pulse shows a behavior
which can be attributed to the properties of a dielectric resonator. This result 
could not have been obtained by the two-dimensional calculation.

c) The amount of energy deposited by the pulse is small and no effect 
observed from exposure of a biological sample to nanopulses can have a 
thermal origin.

Calculation of the electric field surrounding a biological cell is 
a necessary step in understanding effects resulting from exposure to nanopulses.
We have developed a complete FDTD code capable of this. In the near
future, through
the Louisiana Optical Network Initiative, we will have access 
to several supercomputers at 
Louisiana universities connected into one virtual statewide 
supercomputer. We will soon be able to calculate very complicated structures,
larger size objects, and more complex materials.

\section*{Acknowledgments}

I would like to thank Steven P. Wells, Nathan J. Champagne and Arun Jaganathan 
for helpful suggestions.
 
Parts of this material are based on research sponsored by the Air Force 
Research Laboratory, 
under agreement number F49620-02-1-0136. The U.S. Government is authorized to 
reproduce and distribute reprints for Governmental purposes notwithstanding any 
copyright notation thereon. The views and conclusions contained herein are those 
of the authors and should not be interpreted as necessarily representing 
the official policies or endorsements, either expressed or implied, of the 
Air Force Research Laboratory or the U.S. Government.

\section*{References}

\begin{harvard}

\item[] Antar Y M M, Cheng D, Seguin G, Henry B, and Keller M G 1998 
{\it Microwave and Optical Tech. Lett.} {\bf 19} 158
\item[] Balanis C A 1989 {\it Advanced Engineering Electromagnetics} New York:
John Willey \& Sons Inc.
\item[] Dvorak S L and Dudley D G 1995 
{\it IEEE Trans. Electomag. Compatibility} {\bf 37} 192
\item[] Faulkner E A 1969 {\it Introduction to the Theory of Linear Systems} 
London: Chapman and Hall
\item[] Federal Communications Commission 2002, News Release NRET0203, 
http://www.fcc.gov
\item[] Gabriel C 1996 {\it Preprint} AL/OE-TR-1996-0037,
Armstrong Laboratory Brooks AFB, http://www.brooks.af.mil/AFRL/HED/hedr/reports/dielectric/home.html
\item[] Gabriel S, Lau R W  and Gabriel C 1996 {\it Phys. Med. Biol.}  {\bf 41}  2251
\item[] Harrington R F 1961 {\it Time-Harmonic Electromagnetic Fields}
New York: McGraw-Hill
\item[] Harms P H, Lee J F and Mittra R 1992 
{\it IEEE Trans. Microwave Theory Tech.} {\bf MTT-40} 741
\item[] Hu Q, Viswanadham S, Joshi R P, Schoenbach K H, Beebe S J 
and Blackmore P F 2005 {\it  Phys. Rev. E} {\bf 71} 031914
\item[] Jackson J D 1999 {\it Classical Electrodynamics} New York:
John Willey \& Sons Inc.
\item[] Kajfez D and Guillon P Eds. 1986 {\it Dielectric Resonators}
MA: Artech House.  
\item[] Ko W L and Mittra R 1991 
{\it IEEE Trans. Microwave Theory Tech.} {\bf MTT-39} 2176
\item[] Kunz K and Luebbers R 1993 {\it  The Finite Difference Time Domain Method for
Electromagnetics} Boca Raton: CRC Press LLC.
\item[] Luebbers R J, Hunsberger F, Kunz K S, Standler R B, and Schneider M 1990
 {\it IEEE Trans. Electromagn. Compat.} {\bf 32} 222
\item[] Luebbers R J, Hunsberger F, and Kunz K S 1991
{\it IEEE Trans. Antennas Propagat.} {\bf 39} 29
\item[] Luebbers R J and Hunsberger F 1992 {\it IEEE Trans. Antennas Propagat.} 
{\bf 40} 12
\item[] Miller R L, Murphy M R and Merritt J H 2002 
{\it Proceding of the 2nd International Workshop on Biological Effects of EMFs} 
Rhodes Greece
\item[] Mongia R K and Ittipiboon A 1997 
{\it IEEE Trans. Antennas Propagat.} {\bf AP-45} 1348
\item[] Navarro A, Nu\~{n}ez M J and Martin E 1991 
{\it IEEE Trans. Microwave Theory Tech.} {\bf MTT-39} 14
\item[] Pereda J A, Vielva L A, Vegas A and Prieto A 1992 
{\it IEEE Microwave Guided Wave Lett.} {\bf 2} 431
\item[] Polk C and Postow E eds. 1995 {\it Handbook of Biological Effects of
Electromagnetic Fields} Boca Raton: CRC Press LLC.
\item[] Sadiku M N O 1992 {\it  Numerical Techniques in Electromagnetics}
Boca Raton: CRC Press LLC.
\item[] Samn S and Mathur S 1999  {\it Preprint} AFRL-HE-BR-TR-1999-0291,
McKesson HBOC BioServices Brooks AFB.
\item[] Schoenbach K H, Joshi R P, Kolb J F, Chen N, Stacey M, Blackmore P F,
Buescher E S, and Beebe S J 2004 {\it Proceedings of the IEEE} {\bf 92} 1122
\item[] Simicevic N and Haynie D T 2005 {\it Phys. Med. Biol.}  {\bf 50} 347
\item[] Simicevic N 2005 {\it http://caps.phys.latech.edu/$\sim$neven/pulsefield/}
\item[] Sullivan, D M 2000 {\it Electromagnetic Simulation Using the FDTD Method }
New York: Institute of Electrical and Electronics Engineers.
\item[] Taflove A and Brodwin M E 1975{\it IEEE Trans. on
Microwave Theory and Techniques} {\bf 23} 623
\item[] Taflove A and Hagness S C 2000  {\it  Computational Electrodynamics:
The  Finite-Difference  Time-Domain  Method,  2nd  ed.  }  Norwood:  Artech  House.
\item[] Taylor J D ed. 1995 {\it Introduction to Ultra-Wideband Radar Systems}
Boca Raton: CRC Press LLC.
\item[] Van Bladel J 1975 {\it IEEE Trans. Microwave Theory Tech.} {\bf MTT-23} 199;
Van Bladel J 1975 {\it IEEE Trans. Microwave Theory Tech.} {\bf MTT-23} 208
\item[] Vernier P T, Sun Y, Marcu L, Craft C M and Gundersen M A 2004 
{\it Biophys. J.} {\bf 86} 4040
\item[] Yang P, Kattawar G W, Liou K-N and Lu J Q 2004 
{\it Applied Optics} {\bf 43} 4611
\item[] Yee K S 1966 {\it IEEE Trans. Antennas Propagat.} {\bf AP-14} 302

\end{harvard}

\end{document}